\documentclass[12pt]{article}
\usepackage{epsf}

\newcommand{\beq}{\begin{equation}}
\newcommand{\eeq}{\end{equation}}
\newcommand{\beqcol}{\begin{array}{rcl}}
\newcommand{\eeqcol}{\end{array}}

\begin{document}
{\thispagestyle{empty}
\begin{flushright}  YITP-96-64 \\ December 1996 \end{flushright}
\vfill
\begin{center}
{\Large 
Form Factors in $D_n^{(1)}$ Affine Toda Field Theories
}
\\
\vspace{2cm}
{\sc Mathias Pillin{\footnote{Address after February 1st: Department of 
Mathematics, King's College, Strand, London WC2R 2LS, U.K.}} }             \\
\vspace{1cm}
Yukawa Institute for Theoretical Physics  \\ 
Kyoto University                     \\
Kitashirakawa                       \\
606 Kyoto, JAPAN                \\
\                               \\
e-mail: map@yukawa.kyoto-u.ac.jp 
\medskip

\vfill
{\bf Abstract}
\end{center}
\begin{quote}
We derive closed recursion equations for the symmetric polynomials 
occuring in the form factors of $D_n^{(1)}$ affine Toda field 
theories. These equations follow from kinematical- and bound 
state residue equations for the full form factor. We also 
discuss the equations arising from second and third order forward 
channel poles of the S-matrix. The highly symmetric case 
of $D_4^{(1)}$ form factors is treated in detail. We 
calculate explicitly cases with a few particles involved. 
\end{quote}

\eject
}

\setcounter{page}{1}

\section{Introduction}

\bigskip
\bigskip

Form factors are matrix elements of local operators in a 
quantum field theory. Knowing them explicitly or at 
least perturbatively gives a deep knowledge of the 
quantum structure of a given classical Lagrangian 
field theory. For instance it might then in principle 
be possible to classify the local operator content of 
the theory or to calculate correlation functions. 
However, for a generic field theory it is a 
difficult problem to calculate these form factors. 

In the class of two-dimensional field theories one 
is in a better situation. It is known that there 
exist models which enjoy the property that scattering 
is factorizable, i.e. any n-particle scattering 
process can be decomposed into a product of two-particle 
scattering processes. 

For theories whose scattering is factorizable it 
can be shown that the form factors are subject to 
a set of axioms \cite{SMIRNOV}. The axioms (equations) 
provide a machinery which in principle enables one to 
calculate form factors explicitly. 

Within the last years based mainly on the studies 
\cite{KW,SMIRNOV} form factors have been studied 
for some diagonal scattering theories \cite{CM,YZ, ZAM,KM,
FMS1,FMS2,DM,KOUB}. Among them a complete solution to the 
form factor equations was obtained for the 
scaling Lee-Yang-model (minimal $A_2^{(2)}$) \cite{ZAM} 
and for the sinh-Gordon model \cite{KM}. However, 
these models are simple in the sense that they 
do involve only one type of particle or that 
the S-matrices of the corresponding theories 
do only have poles of first order. We note that 
only these first order poles are covered directly by 
the axioms mentioned above \cite{SMIRNOV}. 

Recently the two particle form factors for the magnetic 
perturbation of the Ising model have been investigated 
\cite{DM2}. This work is the first which treats a model 
with several species of particles and an S-matrix with 
higher order poles.

Moreover some form factors in $A_n^{(1)}$ affine Toda 
field theories were studied in \cite{OOTA}. These 
theories do generically have more than one species of 
particles and the S-matrices have poles up to second order. 
Based on the analysis in \cite{OOTA} we would like 
to address the problem of form factors in $D_n^{(1)}$ 
affine Toda field theories. 

\medskip

In order to make this paper a bit more self contained 
let us define affine Toda field theories (ATFT). They 
are given at the classical level by the following 
Lagrangian. 

\beq 
{\cal L} = {1 \over{2}} 
 \partial_{\mu} {\bf \Phi}\cdot \partial^{\mu} {\bf \Phi} 
             - V(  {\bf \Phi} ), \qquad 
V(  {\bf \Phi} ) = {m^2\over{\beta^ 2}} \sum_{i=0}^r n_i 
                   {\rm exp}( \beta \alpha_i \cdot {\bf \Phi})  .
\label{todadef}
\eeq

${\bf \Phi} = ( \Phi_1, \Phi_2, \ldots, \Phi_r)$ is a 
vector of $r$ scalar fields, $\alpha_i$ and $n_i$ denote 
for $i= 1,2,\ldots ,r$ the simple root vectors and the 
Coxeter labels of the corresponding root of 
a simple Lie algebra \cite{KAC}. 
We consider only Lie algebras of type A, D, and E. 

The real constant $m$ sets the mass scale of the theory 
while $\beta$, which is assumed to be real 
throughout this paper, is the coupling of the theory. In 
fact it turns out that as long as we do not consider 
the perturbative structure we can work for our purposes 
with an ``effective coupling'' \cite{BS} 

\beq
B( \beta ) = {1\over{2 \pi}} { \beta^2 \over {1+ \beta^2/ 4\pi}} .
\label{Bdef}
\eeq

The exact S-matrices for ADE-ATFTs have been given in \cite{BCDS1} 
and serve as the main input for the form factor equations. 
It is known that S-matrices for the A-series possess poles 
of at most order two, while for the D-series they do have 
poles up to order four. In the latter case poles of 
first and third order correspond to forward channel 
physical particles. We will come to that point below. 
A general reference on this might be \cite{BCDS2,DOR}. 
The form factor problem for these theories has first been 
addressed in \cite{DDV}. 

In this paper we are going to shed some light on the 
form factors for $D_n^{(1)}$-ATFT. As it was already mentioned 
they provide an interesting complication of the 
studies quoted above since they do contain several 
different particle species and have S-matrices which 
contain higher order poles. 

The outline of the paper is the following. In section 
2 we define form factors and review the axioms which 
they have to obey and some results of \cite{OOTA} which 
will be needed for our study. Some well known facts about 
S-matrices and fusings in $D_n^{(1)}$ are recalled in section 3. 
We then give a closed form of the kinematical 
and bound state residue equations and discuss 
the influence of the higher order poles to the 
form factor equations. In section 4 the case of $D_4^{(1)}$ 
will be discussed in detail. 
We point out how the symmetries of the Dynkin diagram 
enter the form factor equations and analyse in detail 
the poles of second and third order which are present in 
this particular model. We are going to calculate some examples 
of low particle form factors in $D_4^{(1)}$ in section 5. 
The last section is devoted to comments and discussions. 

\bigskip
\bigskip

\section{Some facts about the form factor bootstrap}

\bigskip
\bigskip

As mentioned in the introduction a form factor is a matrix 
element corresponding to a local operator ${\cal O}$. We 
take this matrix element in a special form which is 
commonly used in the literature:

\beq
F_{a_1 \ldots a_n}(\theta_1,\ldots , \theta_n) := 
 \langle 0 | {\cal{O}}(0) | \theta_1,\ldots , \theta_n \rangle .
\label{formdef}
\eeq

In this expression $\theta_i$ labels the rapidity 
of the particle of species $a_i$, and $|0 \rangle$ denotes the physical 
vacuum of the theory. The fact that we take the matrix 
element of ${\cal O}$ at the origin is simply a matter 
of convenience. 

The form factors are known to be subject to four axioms 
\cite{SMIRNOV}. The first two of them are the so called 
Watson's equations. 

\beq
\beqcol
 F_{a_1\ldots  a_i a_{i+1}\ldots  a_n}
(\theta_ 1, \ldots, \theta_ i, \theta_ {i+1}, \ldots, \theta_ n) & =&  
S_{a_ia_{i+1}}(\theta_ i-\theta_ {i+1})F_{a_1\ldots a_{i+1}a_{i}\ldots a_n}
(\theta_ 1, \ldots, \theta_ {i+1}, \theta_ i, \ldots, \theta_ n) ,     \\
& &                                                           \\
 F_{a_1 a_2 \ldots  a_n}(\theta_ 1+2\pi i, \theta_ 2, \ldots, \theta_ n) 
 &=&   F_{a_2 \ldots  a_n a_1}(\theta_ 2, \ldots, \theta_ n, \theta_ 1).
\eeqcol
\label{watson}
\eeq

Needless to mention that $S_{ab}(\theta)$ denotes the S-matrix 
element.

Due to the structure of the theory the form factors do contain 
poles. There are two kinds of them. The first one is due to 
the kinematics. The following equation can be explained by a 
field theoretical reasoning \cite{SMIRNOV} and links a particle 
of species $a$ with its antiparticle $\bar{a}$.

\beq 
-i \ {\rm res}_{\theta'= \theta+ i \pi} \ F_{\bar{a} ad_1 \ldots d_n}
(\theta', \theta,\theta_ 1, \ldots, \theta_ n) = 
F_{d_1\ldots d_n}(\theta_ 1, \ldots, \theta_ n)
\left( 1 - \prod_{j=1}^n S_{ad_j}(\theta-\theta_ j) \right) .
\label{kinres}
\eeq

We will refer to this equation as kinematical residue 
equation. 
 
The second type of singularity is due to the possible 
bound states of the theory. As long as the singularities 
of the S-matrix which lead to bound states are of first 
order the following bound state residue equation for the 
fusion $ a + b \to c$ at fusion angle $\theta^c_{ab}$ and 
with on shell three-point vertex 

\beq 
(\Gamma_{ab}^c)^2 = - i \ {\rm res}_{\theta=i \theta^c_{ab}} S_{ab}(\theta),
\label{vertexdef}
\eeq
holds
\beq
-i \ {\rm res}_{\theta' = \theta+ i \theta_{ab}^c} \ 
F_{a b d_{1}\ldots d_{n}}
(\theta', \theta,\theta_ 1, \ldots, \theta_ n)  =
\Gamma_{ab}^c  F_{\bar{c} d_1 \ldots d_n}
(\theta+ i \bar{\theta}_{bc}^a, \theta_ 1, \ldots, \theta_ n) .
\label{boundres}
\eeq

We have set $\bar{\theta} = \pi - \theta$.

The S-matrix of $D_n^{(1)}$-affine Toda theories is known to 
possess poles of order 2, 3 and 4 as well. The residue equations 
corresponding to these higher order poles need to 
be clarified and will be discussed below.  

\bigskip

The form factors are in general meromorphic functions 
in n variables. It is possible to split the full 
factor into a minimal $F^{ {\rm min}}$ part which is analytic in the strip 
$ 0 \le {\rm Im} \theta< 2 \pi$ and has no zeros 
in $ 0 < {\rm Im} \theta< 2 \pi$. It can be shown 
in generalization of a theorem in \cite{KW} that 
a solution of Watson's equations (\ref{watson}) takes 
the form

\beq
F_{{a_1 \ldots a_n}}  ( \theta_ 1, \ldots \theta_ n) = 
K_{{a_1 \ldots a_n}}  ( \theta_ 1, \ldots \theta_ n) \prod_{i<j} 
F^{{\rm min}}_{a_i a_j} ( \theta_ i- \theta_ j ) ,
\label{split}
\eeq

where it has to hold that $F^{{\rm min}}_{a_i a_j}(\theta) = 
S_{a_i a_j}(\theta) F^{{\rm min}}_{a_j a_i}(-\theta)$ and 
$F^{{\rm min}}_{a_i a_j}(\theta+ 2 \pi i ) = 
F^{{\rm min}}_{a_j a_i}(-\theta)$. These minimal form 
factors are well known for ADE-ATFT. They can be 
presented in an integral form \cite{MAX} or 
equivalently in an infinite product expansion of 
$\Gamma$-functions (see e.g. \cite{OOTA}). 

Now that these are known the work to be done is to 
solve for the form factor equations for the object 
$K$ which contains poles and zeros in the above mentioned 
region. By the construction (\ref{split}) $K$ automatically 
satisfies (\ref{watson}) by requiring some obvious monodromy 
properties. 

It has been shown in \cite{OOTA} that from the kinematical 
(\ref{kinres}) and bound state (\ref{boundres}) residue equations 
we get the following equations for $K$.

\beq
\beqcol
& & -i \ {\rm res}_{\theta^{\prime}=\theta+i\pi} \ 
K_{\bar{a}ad_{1} \ldots d_{n}}
(\theta^{\prime}, \theta,\theta_ {1}, \ldots, \theta_ {n})   =               \\
&=& K_{d_{1}\ldots d_{n}}(\theta_ {1}, \cdots, \theta_ {n})
\left(\prod_{j=1}^{n} \xi_{ad_{j}}(\theta-\theta_ {j})-
\prod_{j=1}^{n} \xi_{ad_{j}}(\theta_ {j}-\theta) \right) /
F^{{\rm{min}}}_{\bar{a}a}(i\pi).
\eeqcol
\label{K-kinemat}
\eeq

Here we have set 

\beq 
\xi_{ab}(\theta)^{-1}= \prod_{x\in A_{ab}} 
\langle x \rangle_{+(\theta)}. 
\label{xi-def}
\eeq

This notation is taken from \cite{OOTA} and will be explained 
in the beginning of the next section (\ref{notation1}). 
Here we would only like 
to state that the S-matrix is given by 

\beq
S_{ab}(\theta) = { {\xi_{ab}(-\theta)}\over{\xi_{ab}(\theta)}} .
\label{S-mat-xi}
\eeq
This form should be compared with the expressions for the 
S-matrix at the beginning of the next section.

The bound state residue equation is

\beq
\beqcol
& & -i \ {\rm res}_{\theta^{\prime} = \theta+ i \theta_{ab}^c} \ 
K_{abd_{1}\ldots d_{n}} (\theta^{\prime}, \theta, \theta_ 1, \cdots,
\theta_ n)               =             \\
&=& \Gamma_{ab}^c K_{\bar{c} d_1 \ldots d_n}
(\theta+i\bar{\theta}_{bc}^a, \theta_ 1, \cdots, \theta_ n)
\prod_{j=1}^n  \lambda_{ab;d_j}^c (\theta+i\bar{\theta}_{bc}^a-\theta_ j)
/ F^{{\rm{min}}}_{ab}(i\theta_{ab}^c).
\eeqcol
\label{K-bound}
\eeq

The object $\lambda$ is via Watson's equation very closely 
related to the S-matrix bootstrap equation. Since its 
explicit form will be of importance later we mention:

\beq
\lambda^c_{ab;d}(\theta)^{-1} = 
{ F^{{\rm min}}_{ad} (\theta+ i \bar{\theta}^b_{ac} ) 
  F^{{\rm min}}_{bd}(\theta- i \bar{\theta}^a_{bc} ) \over 
{F^{{\rm min}}_{\bar{c} d} (\theta) }} = 
\prod_{x\in A_{ad} \atop{x\le \bar{u}^b_{ac}}} 
      \langle \bar{u}^b_{ac}- x \rangle_{+(\theta)} 
\prod_{x\in A_{bd} \atop {x < \bar{u}^a_{bc}}} 
      \langle x-\bar{u}^a_{bc} \rangle_{+(\theta)} .
\label{lamdef}
\eeq

The $\theta^{a}_{bc}$'s ($\bar{\theta}^{a}_{bc} = \pi - \theta^{a}_{bc}$) 
are the fusion angles of the theory and 
we set $u^{a}_{bc}= \theta^{a}_{bc} h/ \pi $, 
with $h$ being the Coxeter number.    

We will give details on this notation at the beginning of 
the next section.

\bigskip

We have now reduced the problem of calculating form 
factors in ATFT to that of making a proper ansatz for $K$ and 
solving the two equations (\ref{K-kinemat}) and 
(\ref{K-bound}). In cases where the S-matrix has 
higher order poles we do have to supplement these 
equations with additional ones. 

\bigskip
\bigskip

\section{ Form Factors in $D_n^{(1)}$ Toda theory}

\bigskip
\bigskip

Let us recall some facts about $D_n^{(1)}$-ATFT \cite{BCDS1}. 
These theories contain n kinds of particles which 
will be labeled by the elements of the set $\{ 1,2,\ldots ,n-2, s, 
s^{\prime}\}$. While the particles labeled by ordinary 
numbers in this set correspond to the points on the 
straight line, $s$ and $s^{\prime}$ correspond to the 
particle upside and downside of the fishtail part 
of the Dynkin diagram of $D_n$. 

In this class of ATFT we have to distinguish between $n$ odd 
and $n$ even. In the even case all particles are 
self-conjugate while in the odd case we have $\bar{s} = s^{\prime}$. 

Let $ h = 2(n-1)$ denote the Coxeter number of the theory 
in question. The masses of the particles are then given by

\beq
m^2_s = m^2_{s^{\prime}} = 2 m^2, \qquad m^2_k = 8 m^2\  
 {\rm sin}^2 ({{k \pi}\over{h}}), \qquad k=1,2,\ldots, n-2 .
\eeq
$m^2$ sets the mass scale of the theory. We remark that the 
case of $D_4^{(1)}$ is special in the sense that here we 
do have three particles $\{ 1,s,s^{\prime} \}$ of mass $\sqrt{2}m$ and 
one heavy particle $\{ 2 \}$ of mass $\sqrt{6}m$ which corresponds 
to the central point of the Dynkin diagram of $D_4$.  

\medskip

The exact S-matrices and the fusion angles for the 
$D_n^{(1)}$-theories have been given in \cite{BCDS1}. We recall them for
convenience. In doing that we borrow the following notations 
from \cite{OOTA}.

\beq
(r)_{+(\theta)} := {1\over{i \pi}} \ {\rm sinh} {1\over{2}} 
\left(\theta+ {{i\pi} \over{h}} r \right), \qquad 
(r)_{(\theta )} := { (r)_{+(\theta)} \over {(-r)_{+(\theta)}}}, 
\eeq
and
\beq
\langle r \rangle_{+(\theta)}= {{(r+1)_{+(\theta)}(r-1)_{+(\theta)}} 
\over {(r+1-B)_{+(\theta)}(r-1+B)_{+(\theta)}}}, \qquad 
\langle r \rangle_{(\theta)}= { {\langle r \rangle_{+(\theta)}} \over 
{\langle -r \rangle_{+(\theta)}}} .
\label{notation1}
\eeq

Using these notations the S-matrices for n even are given by
\beq
S_{ab}(\theta) = \prod_{{|a-b|+1 \atop {\rm  step} 2}}^{a+b-1} 
 \langle p \rangle_{(\theta)}  \langle h-p \rangle_{(\theta)}, \qquad
S_{sa}(\theta) = S_{s^{\prime}a}(\theta) = 
\prod_{0\atop {\rm step} 2}^{2a-2} \langle n-a+p \rangle_{(\theta)}  ,
\eeq
with $a,b \in {1,2,\ldots, n-2}$, and 
\beq 
S_{ss}(\theta)=S_{s^{\prime}s^{\prime} }(\theta) = 
\prod_{1 \atop {\rm step} 4}^{h-1} \langle p \rangle_{(\theta)} , \qquad
S_{s s^{\prime}}(\theta) = \prod_{3 \atop {\rm step} 4}^{h-3} 
\langle p \rangle_{(\theta) } 
\eeq

For odd n similar expressions are obtained \cite{BCDS1}.

We denote by $A_{ab}$ the set of integers actually appearing 
in the brackets of the corresponding S-matrix above and by 
$\hat{A}_{ab}= A_{ab} \setminus \{ h-1 \}$. For later 
calculations it turns out to be useful to know 
the number of elements in these sets. 

\beq
\beqcol 
\# A_{aa} &=& 2 a,    \\
\# A_{ab} &=& 2 a,    \\
\# A_{as} &=& a,       \\
\# A_{ss} &=& \# A_{s^{\prime} s^{\prime}} = n/2,  \\
\# A_{s s^{\prime}} &=&  (n-2)/2, 
\eeqcol
\qquad
\beqcol
h-1 & \in & A_{aa},            \\
h-1 & \in {\hspace{-0.3cm}} / & A_{ab} ,\quad a < b,    \\
h-1 &\in {\hspace{-0.3cm}} /  &A_{as} ,    \\
h-1 & \in & A_{ss}, A_{s^{\prime} s^{\prime}} , \\
h-1 & \in {\hspace{-0.3cm}} / & A_{s s^{\prime}} .
\eeqcol
\label{A-number}
\eeq
 
We assumed $a\ne b \in {\hspace{-0.3cm}} /  \{ s,s^{\prime} \} $. 
For odd n these results have to be slightly modified.

\medskip

The fusion angles can be divided into three classes. The 
first one involves the particles $s$ and $s^{\prime}$. 

\beq 
u_{ss}^a= u_{ss^{\prime}}^a = h-2a, \qquad 
u_{sa}^s= u_{s^{\prime}a}^s = h/2 +a .
\label{s-fusion}
\eeq

The second and third class do not incorporate $s$ and $s^{\prime}$ 
and are distinguished by a relation of the three particles 
in the the fusion process. We have
\beq 
\beqcol
u_{ab}^c &=& h-c, \quad u_{ac}^b = h-b, \quad 
u_{bc}^a = h-a, \qquad {\rm if} \quad a+b+c=h,    \\
& &                                                      \\
u_{ab}^c& =& h-c, \quad u_{bc}^a=h-a, \quad u_{ac}^b=b, 
\qquad \quad \ {\rm if} \quad a-b+c=0.
\eeqcol
\label{a-fusion}
\eeq

\bigskip

Up to now we had to review a lot of known things and to 
introduce a bulk of notation. We are now in a position 
to come to the main part of this paper. 

\bigskip

There might be several ways to write down the part 
of the full form factor which contains poles 
and zeros in the physical region. We prefer a 
slightly modified version of what has been 
presented in \cite{OOTA}. This version is based 
on a factorization of the pole part of $K$ while 
the part containing zeroes is the quantity to be 
determined by the form factor equations. We 
take $x_i= e^{\theta_ i}$ and define 
the follwing vectors

\beq 
{\bf x}^{(k)} = ( x^{(k)}_1, x^{(k)}_2,\ldots , x^{(k)}_{N_k} ), 
\qquad k \in \{ 1,2, \ldots , n-2,s, s^{\prime} \}.
\eeq
The $N_k$ denote the number of particles of type $k$ 
in the form factor (cf. (\ref{formdef})). 
We can then write down the following parametrization:

\beq
\beqcol
  K_{[N_1,\ldots , N_{n-2},N_s,N_{s^{\prime}}]} 
( {\bf x}^{(1)}, \ldots , {\bf x}^{(n-2)},{\bf x}^{(s)},
{\bf x}^{(s^{\prime})} ) & = &  
Q_{[N_1,\ldots , N_{s^{\prime}}]} 
( {\bf x}^{(1)}, \ldots ,{\bf x}^{(s^{\prime})} )  \\ 
& &                                                \\
& &  {\hspace{-6cm}} \times 
 ( \prod_{k=1}^{s^{\prime}} \prod_{i<j}^{N_k}  
{ 1 \over{x^{(k)}_i+ x^{(k)}_j} } 
{ 1\over{ W_{k k}(x^{(k)}_i , x^{(k)}_j ) }} )
\prod_{k=1}^{s} \prod_{l=k+1}^{s^{\prime}} 
\prod_{i=1}^{N_k} \prod_{j=1}^{N_l} 
 { 1\over{ W_{k l}(x^{(k)}_i , x^{(l)}_j ) }}     .
\eeqcol
\label{K-ansatz}
\eeq

The factors $1/( x^{(k)}_i+ x^{(k)}_j )$ parametrize 
the kinematical singularities. The product in front of 
them is already chosen for $D_n^{(1)}$ with n even, i.e. 
all particles are self conjugate. Obviously we can 
write down $K$ without any difficulties for 
odd n as well. In what follows we give the 
explicit formulas for the cases with even n only.

The objects $ W_{k l}(x^{(k)}_i , x^{(l)}_j ) $ do 
parametrize the fusion poles. In \cite{OOTA} it was 
shown that a particular ansatz for them does 
lead from (\ref{K-kinemat}) and (\ref{K-bound}) to 
a polynomial equation for the $Q$'s at least for 
$A_n^{(1)}$-ATFT. We take this ansatz in a slightly 
modified form. 

\beq
W_{k l}(x^{(k)}_i , x^{(l)}_j ) =  \prod_{r \in \hat{A}_{kl} }
( x^{(k)}_i - \Omega^{r+1} x^{(l)}_j ) \ 
( x^{(k)}_i - \Omega^{-r-1} x^{(l)}_j ) , \qquad \Omega= e^{i \pi /h}.
\label{W-def}
\eeq

In the ansatz (\ref{K-ansatz}) above 
$Q_{[N_1,N_2,\ldots , N_{s^{\prime}}]}$ is 
assumed to be a polynomial which is due to the structure 
of ATFT symmetric at least in each of the vector 
components of any ${\bf x}^{(k)}$. Of course this 
assumption has to be justified case by case especially 
in the presence of higher order poles in 
ATFT. Since our form factors are required to be 
Lorentz invariant it is straightforward to 
calculate the degree of the polynomials $Q$. A 
Lorentz transformation consists simply of a linear 
shift of the rapidities. The denominator of (\ref{K-ansatz}) 
is not Lorentz invariant. However its degree determines 
the total degree of $Q$ to be:

\beq
\beqcol
{\rm deg}\  Q_{[ N_1,\ldots , N_{s^{\prime}}]} & = & 
\sum_{k=1}^{n-2} \left( N_k( N_k -1) ( 2 k - {1\over 2} ) +  
2 N_k (N_s+ N_{s^{\prime}} ) k  \right) + 
4  \sum_{k=1}^{n-3} \sum_{l=k+1}^{n-2} k N_k N_l           \\
& &                                                      \\
& + & {{n-1}\over{ 2}} \sum_{k=s,s^{\prime}} N_k ( N_k -1) + 
N_s N_{s^{\prime}} ( n-2)  .
\eeqcol
\label{Q-degree}
\eeq
 
The partial degree, i.e. the maximal degree of a variable 
$x_i^{(k)}$ in $Q$ can be calculated only after having 
established the recursion equations for the polynomials 
below. In general this leads to a quite complicated 
formula. We treat the problem of calculating the 
partial degree for the special case of $D_4^{(1)}$ in 
section 5.

\medskip

It is now clear that we have chosen an ansatz for the 
form factor in such a way that we have to solve only for 
the polynomials $Q$. However, as can be read off from 
(\ref{Q-degree}) 
these objects do have a quite high degree already in cases 
when only a few particles are present. This makes the 
general solution technically quite difficult. 

\bigskip

We can now plug the above ansatz into the residue 
equations to obtain recursion relations for the polynomials 
$Q$. In doing this we first make the following observation

\beq
\beqcol
K_{ab\ [ N_1,\ldots , N_{s^{\prime}} ]} ( x^{(a)}, x^{(b)}, 
{\bf x}^{(1)}, \ldots ,{\bf x}^{(s^{\prime})} ) = 
(x^{(a)}+x^{(b)})^{-\delta_{ab}}     
\prod_{i=1}^{N_{a}}(x^{(a)}+x^{(a)}_i )^{-1} 
\prod_{i=1}^{N_{b}}(x^{(b)}+x^{(b)}_i )^{-1} \                \\
  & &                                                \\
 & & {\hspace{-15cm}} \times
\prod_{k=1}^{s^{\prime}} \prod_{i=1}^{N_k} W_{ak}(x^{(a)},x^{(k)}_i)^{-1} 
W_{bk}(x^{(b)},x^{(k)}_i)^{-1} \cdot W_{ab}(x^{(a)},x^{(b)})^{-1} \\
& &                                                    \\ 
& & {\hspace{-15cm}}\times Q_{ab\ [N_1,\ldots , N_{s^{\prime}}]} 
(x^{(a)}, x^{(b)}, {\bf x}^{(1)}, \ldots {\bf x}^{(s^{\prime})} )  \ \ 
{ K_{[N_1,\ldots , N_{s^{\prime}}]} \over  
Q_{[N_1,\ldots , N_{s^{\prime}}]}}
({ \bf x}^{(1)}, \ldots ,{\bf x}^{(s^{\prime})} ).
\eeqcol
\label{ab-factor}
\eeq
This equation indicates how the two particles $a$ and $b$ 
undergoing either a kinematical or a bound state process 
are separated from all other particles in the 
form factor. The last factor in the product does 
not contain any dependence on the particles $a$ and $b$, 
i.e. the coordinates $x^{(a)}$ and $x^{(b)}$ do not appear 
in that expression. A similar expression can of course 
be written down if we want to seperate only one particle.  
   
By using the structure of the index sets $A_{ab}$ of 
the exact S-matrices for $D_n^{(1)}$-ATFT and 
introducing the following piece of notation to 
make expressions a bit more transparent

\beq
[r]^{(k)}_i = x- \Omega^r \ x^{(k)}_i, \qquad k \in \{ 1,2, \ldots, n-2,s,
s^{\prime} \}, \quad i = 1,2, \ldots N_k ,
\label{brack-def}
\eeq

we arrive 
from the kinematical residue equation (\ref{K-kinemat}) 
at the following recursion equation which links an 
$N+2$-particle polynomial to an $N$-particle polynomial.

\beq
\beqcol
& & Q_{aa [N_1 \ldots N_{s^{\prime}}] } 
( -x^{(a)}, x^{(a)}, {\bf x}^{(1)}, 
      {\bf x}^{(2)},\ldots , {\bf x}^{(s^{\prime})} ) = 
-i (-1)^{N_a} (x^{(a)})^{4a-1} 
 { {\prod_{r\in {\hat{A}}_{aa}} 2 {\rm cos}( {{(r+1) \pi}\over{2h}} )} \over 
   {F^{{\rm{min}}}_{aa}(i \pi)}}                                   \\
& &                                                           \\
     &\times & ( 
   \prod_{k=1}^{s^{\prime}} \prod_{i=1}^{N_k} 
   \prod_{r\in{\hat{A}}_{ak}} [ -r-1+h ]_i^{(k)} [r+1 ]_i^{(k)}
   \prod_{r\in A_{ak}}  [ 1-r-B  ]_i^{(k)} [ B-r-1 ]_i^{(k)}    \\
& &                                                               \\ 
   &\ \ & - \prod_{k=1}^{s^{\prime}}\prod_{i=1}^{N_k} 
   \prod_{r\in{\hat{A}}_{ak}} [ -h+r+1 ]_i^{(k)} [ -r-1  ]_i^{(k)}
   \prod_{r\in A_{ak}} [ r-1+B ]_i^{(k)} [ r+1-B  ]_i^{(k)} )     \\
& &                                                               \\
   & \times & Q_{[N_1 \ldots N_{s^{\prime}}] } 
( {\bf x}^{(1)},  {\bf x}^{(2)},\ldots , {\bf x}^{(s^{\prime})} ).
\eeqcol
\label{Q-kinemat}
\eeq

Obtaining a closed recursion relations for the polynomials 
$Q$ coming from the first order bound state equation 
(\ref{K-bound}) is not so straightforward. It was mentioned 
above that the fusings in $D_n^{(1)}$-ATFT can be devided 
into three classes (\ref{s-fusion}), (\ref{a-fusion}). In 
the second class of (\ref{a-fusion}) the three particles 
entering appear to be unsymmetric in the sense $a-b+c=0$. 
In order to calculate the desired recursion relation 
these three classes have to be treated seperately and 
especially the third class carefully. We also have to take 
into account the number of elements in the sets $A_{ab}$ 
(\ref{A-number}). The first order equation for 
the process $a+b \to c$ is then

\beq
\beqcol
Q_{ab [N_1,\ldots , N_s^{\prime}]} ( x \Omega^{\Theta^{c}_{ab}} , 
x \Omega^{-\bar{u}^{a}_{bc}},  {\bf x}^{(1)},\ldots , {\bf x}^{(s^{\prime})} )
 &=& { { \Gamma^{c}_{ab} } \over {F^{\rm{min}}_{ab} ( i \theta^c_{ab} ) 
     (-i) {\rm{res}}_{ x^{\prime}=x \Omega^{\Theta^{c}_{ab}}} 
   W_{ab}(x^{\prime},x )^{-1} }}                                    \\
 & & {\hspace{-9cm} } 
\Omega^{ {\Theta^{c}_{ab}} (N_a + \sum_{k=1}^{s^{\prime}} 
    N_k \# \hat{A}_{ak} ) - \bar{u}^{a}_{bc} ( N_b + 
    \sum_{k=1}^{s^{\prime}} N_k \# \hat{A}_{bk} ) } 
\prod_{i=1}^{N_a} ([u^b_{ac}]^{(a)}_i)^{-1} 
\prod_{i=1}^{N_b} ([-u^a_{bc}]^{(b)}_i)^{-1} 
\prod_{i=1}^{N_c} [h]^{(c)}_i                               \\
& &                                                        \\
& & {\hspace{-9cm}} \times
\prod_{k=1}^{s^{\prime}} \prod_{i=1}^{N_k} (
\prod_{r\in A_{ak}\atop r \le \bar{u}^b_{ac} }
[1+r+u^b_{ac}]^{(k)}_i [-1+r+u^b_{ac}]^{(k)}_i  
[-\bar{u}^b_{ac}+r-B+1]^{(k)}_i  [-\bar{u}^b_{ac}+r+B-1]^{(k)}_i \\
& &                                                          \\
& & {\hspace{-9cm}} \times 
\prod_{r\in A_{bk}\atop r < \bar{u}^a_{bc} } 
[-u^a_{bc}-r+1]^{(k)}_i [-u^a_{bc}-r -1]^{(k)}_i  
[-r+\bar{u}^a_{bc}+B-1]^{(k)}_i  [-r+\bar{u}^a_{bc}-B+1]^{(k)}_i )   \\
& &                                                               \\
& \times &  Q_{c[N_1 \ldots N_{s^{\prime}}] } 
(x, {\bf x}^{(1)}, \ldots , {\bf x}^{(s^{\prime})} ) .
\eeqcol
\label{Q-bound}
\eeq

The last two lines should be compared with the explicit 
expression for $\lambda$ given in (\ref{lamdef}). 

$u^{c}_{ab}$ etc. are the fusion angles for the 
possible processes which involve the particles 
$a$, $b$, and $c$. We used the abbreviation 
$\Theta^{c}_{ab}= u^c_{ab} + u^a_{bc} -h$. For 
self-conjugate theories it holds of course 
that $\Theta^{c}_{ab}=\bar{u}^b_{ac}$.

Even though we do have negative powers in the 
above expression the equation is a polynomial 
recursion equation. It can be checked that 
these negative powers are canceled for any ATFT. We 
have, however, chosen to write them in order to 
make the remaining products a bit more handy.

We note already at this stage that the recursion 
coefficients in (\ref{Q-kinemat}) and (\ref{Q-bound}) can 
be expressed as a sum of elementary symmetric polynomials. 
We will treat this point more explicitly in section 5 
when we calculate some solutions to these equations 
for the $D_4^{(1)}$ case. 

In order to be able to obtain solutions to these 
equations without going into tedious case by case 
calculations it would be desirable to have a 
Lie algebraic interpretation of the two recursion 
equations for instance in the sense of \cite{DOR}. 
We were informed \cite{OOTA2} that an equation similar 
to (\ref{Q-bound}) can also be obtained for the $Q$-polynomials 
arising in $A_n^{(1)}$-form factors. 

It is important to note that (\ref{Q-bound}) is written 
if the particles are in order. This means that we impose 
an ordering in the set of particles $\{ 1,2, \ldots n-2 , 
s,s^{\prime} \}$ by $ 1 <2, \ldots 1< s^{\prime}, \ldots , 
s < s^{\prime} $. Then in (\ref{Q-bound}) we have $a \le b$. 

If we change this order of the particles in the process, i.e. we consider 
$b+a \to c$ rather than $a+b\to c$ the coefficient 
in (\ref{Q-bound}) will change. This change is simply 
obtained by exchanging the first two arguments of 
$Q$ and replacing $\Omega \to \Omega^{-1}$ in the coefficient. 
Since both processes are physically equivalent 
this artifact leads to an identity for one 
particular polynomial $Q$ at specific values 
of the arguments. This will become 
more transparent in section 5.

\bigskip

Let us now come to a discussion of the form factor 
equations in the presence of higher order poles of 
the S-matrix. The content of equation (\ref{Q-bound}) can 
be stated graphically by the following diagram which should 
not strictly be intrepreted in the perturbative sense. 

\bigskip
\bigskip

{\hspace{5cm}}
{\input{figure1.eps}}

\bigskip
\bigskip

The bubble in this figure depicts the local operator 
in the conventions of (\ref{formdef}).

\medskip

Let us consider the third order foward channel 
poles of the S-matrix first. These poles correspond 
to fusings which are already covered by the 
fusion angle analysis at the beginning of 
this section. For our purposes this means the 
following. The perturbative analysis shows that in 
general several Feynman diagrams are needed in order 
to explain the third order poles of the S-matrix. 
But the S-matrix in our case is known to satisfy 
the bootstrap equations. Hence, for the forward channel 
fusings at the position of a third order pole of 
the S-matrix we can in analogy to fig.1 draw the 
following effective picture.

\bigskip
\bigskip

{\hspace{5cm}}
{\input{figure2.eps}}

\bigskip
\bigskip

The circle at the position of the fusing 
indicates that we do not need to know 
about the perturbative processes which lead to the 
production of the physical particle running into 
the local operator. 

\medskip

This means that we should be able to derive equations 
for the $Q$-polynomials corresponding to fusings of 
the type in fig.2 in analogy to what has been done 
for the first order pole fusings. 

A detailed analysis shows that this is can consistenly 
be achieved. The 
result for a fusing $a+b \to c$ at a third order 
pole of the S-matrix is exactly equation 
(\ref{Q-bound}) where we only have to replace the 
residue appearing in the denominator of the right 
hand side of (\ref{Q-bound}) by the coefficient 
of the leading order singularity of 
$W_{ab}(x^{\prime},x)^{-1}$.

In other words, due to the bootstrap properties of the 
S-matrix the fusings corresponding to third order 
poles of the S-matrix can be treated on exactly the 
same footing like the first order ones and the 
corresponding recursion equations for the $Q$-polynomials 
are structurally identical!

We conjecture that the same phenomenon is true for 
the higher order forward channel fusings in the 
ATFTs coming from the E-series as well. 

The problem of deriving equations for the form factor in 
presence of higher odd order poles of the S-matrix has 
already been addressed in the context of Ising models 
in \cite{DM2}.

\bigskip
\bigskip
 
The question arises what happens in the case of the 
second and fourth order poles of the S-matrix. For 
a discussion of this question see also \cite{DM2}. These 
poles are known \cite{BCDS1} not to participate in the 
S-matrix bootstrap. It is therefore not possible 
to draw effective diagrams like the one in fig. 2 for these 
cases. One might therefore be tempted to think that 
these poles do not lead to any significant 
contribution on the form factor level. 
However, this is not true. We will see in the next 
section that it is possible to derive a 
consistent equation for the $Q$-polynomials corresponding 
to the second order pole processes of the S-matrix 
in $D_4^{(1)}$-ATFT. Moreover, it will turn out in 
section 5 that these equations are needed in order 
to constrain the solution spaces of the $Q$-polynomials.

Hence it seems that for the even order poles the processes 
which are needed in order to explain the corresponding 
S-matrix element have an influence on the solution 
spaces of the form factors. It is, however, not clear 
how to derive in general equations for the $Q$-polynomials 
in these cases even though it might in principle be clear how 
the equations for the full form factors can be formulated 
(see e.g. \cite{DM2,OOTA,MUSS} and (\ref{second-order})). 
The main reason for this mainly the fact that in 
$D_n^{(1)}$-ATFT (in contrast to the $A_n^{(1)}$ cases) 
we do not know a priori how many Feynman diagrams do 
contribute to an even order pole of the S-matrix and what 
kinds of particles are involved in a particular 
diagram. Therefore, up to our present knowledge the 
contributions of these diagrams have to be studied 
case by case. As mentioned above we are going to treat 
the case of the second order pole in $D_4^{(1)}$ in 
detail in the following sections.

\bigskip
\bigskip

\section{The $D_4^{(1)}$ case}

\bigskip
\bigskip

We are now going to discuss the form factors in 
the simplest $D_n^{(1)}$-ATFT. As mentioned above 
the $D_4^{(1)}$-ATFT describes four particles 
which we label $1$, $2$, $s$, and $s^{\prime}$. The 
mass of particle $2$ is $m_h = \sqrt{6} m$ while 
the masses of the other particles are the same 
with $m_l= \sqrt{2} m$. This means that we 
can devide the particle content into light 
particles $l \in \{ 1,s,s^{\prime} \}$ and one heavy 
particle $h=2$. This kind of symmetry follows 
from the symmetry of the Dynkin diagram of $D_4$ 
(see e.g. \cite{KAC}). In this section we are going to present the 
form factor equations of this particular model. 
We will see that the symmetries of the Dynkin diagram 
are maintained at the level of the form factors. 

Even though this model is quite simple compared to 
the general $D_n^{(1)}$-models it already shows 
the generic features which are present in 
$D_n^{(1)}$-theories. Referring to the S-matrix 
we do have a third order 
fusion pole for the process $h+h \to h$ and a 
second order pole for the process $l+h \to l^{\prime} 
+ l^{\prime\prime}$. However, the S-matrix does 
not have fourth order poles which makes life a 
bit easier. 

\medskip

We are going to apply the machinery presented in the 
previous section. First we use (\ref{Q-degree}) and 
obtain for the total degree of our $Q$-polynomials:

\beq
\beqcol
{\rm deg}Q_{[N_1,N_2,N_s,N_{s^{\prime}}]} &=&  
 {3\over 2} \left(  
        N_1(N_1-1) + N_s(N_s-1) + N_{s^{\prime}}(N_{s^{\prime}}-1)
\right) + {7\over 2} N_2(N_2-1)                                    \\
& &                                                               \\
&+& 2\left( N_1 N_s + N_1 N_{s^{\prime}} 
+ N_s N_{s^{\prime}} \right) + 4 N_2 \left( N_1 + N_s 
+N_{s^{\prime}} \right) .
\eeqcol
\label{Q4-degree}
\eeq

Using the structure of the S-matrices we obtain the 
following equations for the kinematical poles. 

\beq
\beqcol
Q_{ll\ [N_1,N_2,N_s,N_{s^{\prime}}] } ( -x,x,
{\bf{x}}^{(1)},\ldots , {\bf{x}}^{(s^{\prime})} ) & = & 
x^3 { { - 3 i \ (-1)^{N_l}}\over {F_{ll}^{\rm{min}}(i \pi ) }}  \\ 
& &                                                             \\
 & & {\hspace{-6cm}} 
( \prod_{i=1}^{N_l} [4]_i^{(l)} [2]_i^{(l)} [-B]_i^{(l)} 
[B-2]_i^{(l)} [-B-4]_i^{(l)} [B-6]_i^{(l)}\quad                    \\
& &                                                               \\
 & & {\hspace{-6cm}} 
\prod_{j=1}^{N_h}  ([3]_j^{(h)})^2  [1]_j^{(h)}  [5]_j^{(h)}  
[-B-1]_j^{(h)}  [B-3]_j^{(h)}  [-B-3]_j^{(h)}  [B-5]_j^{(h)}    \\ 
 & &                                                              \\
 & & {\hspace{-6cm}} 
\prod_{k=l^{\prime}, l^{\prime\prime}} \prod_{m=1}^{N_k} 
  [2]_m^{(k)} [4]_m^{(k)} [-B-2]_m^{(k)} [B-4]_m^{(k)}           \\
& &                                                               \\
& & {\hspace{-4cm}} -  {\hspace{2cm}}[r] \leftrightarrow [-r] ) \ 
Q_{[N_1,N_2,N_s,N_{s^{\prime}}] } 
({\bf{x}}^{(1)},\ldots , {\bf{x}}^{(s^{\prime})} ) .
\eeqcol
\label{ll-kinemat}
\eeq  

The notation in the last line indicates that on the right 
side of the minus sign appears a polynomial structurally 
identical the the one on the left side with all the entries 
in the brackets replaced by their negative values.  

In this expression $l$ can be either $1$, $s$, or $s^{\prime}$ 
and therefore shows the symmetry of the theory on the 
form factor equation level. 

For the kinematical pole of particle $h$ we find

\beq
\beqcol
Q_{hh\ [N_1,N_2,N_s,N_{s^{\prime}}] } ( -x,x,
{\bf{x}}^{(1)},\ldots , {\bf{x}}^{(s^{\prime})} ) & = & 
x^7 { { - 3 i \ (-1)^{N_h}}\over {F_{hh}^{\rm{min}}(i \pi ) }}  \\ 
& &                                                             \\
 & & {\hspace{-7cm}} 
( \prod_{k=l,l^{\prime}, l^{\prime\prime}} 
 \prod_{i=1}^{N_k} [1]_i^{(k)} ([3]_i^{(k)})^2 [5]_i^{(k)} 
 [-B-1]_i^{(k)} [B-3]_i^{(k)} [-B-3]_i^{(k)} [B-5]_i^{(k)}\quad     \\
& &                                                               \\
 & & {\hspace{-7cm}} 
\prod_{j=1}^{N_h}  ([2]_j^{(h)})^3 ([4]_j^{(h)})^3 [-B]_j^{(h)}  
([-B-2]_j^{(h)})^2  [-B-4]_j^{(h)}  [B-2]_j^{(h)}  
([B-4]_j^{(h)})^2 [B-6]_j^{(h)}    \\ 
 & &                                                              \\
 & & {\hspace{-4cm}}- {\hspace{2cm}}[r] \leftrightarrow [-r] ) \ 
Q_{[N_1,N_2,N_s,N_{s^{\prime}}] } ({\bf{x}}^{(1)},\ldots , 
{\bf{x}}^{(s^{\prime})} ) .
\eeqcol
\label{hh-kinemat}
\eeq

\bigskip
\bigskip

To get the recursion equations for 
the first order bound-state residue equations we introduce 
the following notation to make the formulae as transparent as possible.
\beq
\{ r \}_i^{(k)}= [ r+1 ]_i^{(k)} [ r-1 ]_i^{(k)} 
             [ -h+r+1-B  ]_i^{(k)} [ -h+r-1+B ]_i^{(k)} .
\label{brack3-def}
\eeq
Here $h$ stands for the Coxeter number which is equal to 
six in the present case.

\bigskip
\bigskip
  
We can derive the following recursion relations in $D_4^{(1)}$ where 
we put some constant factors appearing in (\ref{Q-bound}) into a 
constant $H$ for the particular process. 

First we get for  $l+l \to h$:
\beq 
\beqcol
 Q_{ll [N_1 \ldots N_{s^{\prime}}] } 
( x\Omega , x\Omega^{-1} , {\bf x}^{(1)}, 
      {\bf x}^{(2)},{\bf x}^{(s)}, {\bf x}^{(s^{\prime})} ) & = &  
{\hspace{4cm}}\\
& &                                                            \\
& & {\hspace{-7cm}} H_{ll} \ 
  x^3 \ \prod^{N_l}_{i=1} { { \{ 6 \}_i^{(l)}}\over{ [ 5 ]_i^{(l)} 
[ 7 ]_i^{(l)}}} \prod^{N_h}_{i=1} [ 6 ]_i^{(h)}  \ 
Q_{ h[N_1 \ldots N_{s^{\prime}}] } 
( x , {\bf x}^{(1)}, 
      {\bf x}^{(2)},{\bf x}^{(s)}, {\bf x}^{(s^{\prime})} ) 
\eeqcol
\label{ll-bound}
\eeq

The next processes to be considered are $l+h \to l$ 
and $l+l^{\prime} \to l^{\prime\prime}$. Using (\ref{Q-bound}) 
we arrive at the following recursion relations 

\beq
\beqcol
 Q_{lh [N_1 \ldots N_{s^{\prime}}] } 
( x\Omega^4 , x\Omega^{-1} , {\bf x}^{(1)},\ldots , 
{\bf x}^{(s^{\prime})} ) & = & H_{lh} \ x^4 \  
\Omega^{6 N_l + 4 N_h + 2 ( N_{l^{\prime}} + N_{l^{\prime\prime}})}  \\
& &                                                            \\
& & {\hspace{-7cm}}  \times
\prod^{N_l}_{i=1} { { \{ 3 \}_i^{(l)} [6]_i^{(l)}}
\over{ [ 4]_i^{(l)}}} 
\prod^{N_h}_{i=1} { { \{ 4 \}_i^{(h)} \{ 6 \}_i^{(h)}} 
\over {[7]_i^{(h)}}} 
\prod^{N_{l^{\prime}}}_{i=1} \{ 5 \}_i^{(l^{\prime})}
\prod^{N_{l^{\prime\prime}}}_{i=1} \{ 5 \}_i^{(l^{\prime\prime})}  
Q_{ l[N_1 \ldots N_{s^{\prime}}] } 
( x , {\bf x}^{(1)},\ldots , {\bf x}^{(s^{\prime})} ) ,
\eeqcol
\label{lh-bound}
\eeq

and

\beq
\beqcol 
Q_{ll^{\prime} [N_1 \ldots N_{s^{\prime}}] } 
( x\Omega^2 , x\Omega^{-2} , {\bf x}^{(1)},\ldots , 
{\bf x}^{(s^{\prime})} ) & = & H_{ll^{\prime}} \ x^2 \  
\Omega^{2 ( N_l - N_{l^{\prime}} ) }                            \\
& &                                                            \\
& & {\hspace{-7cm}}  \times
\prod^{N_l}_{i=1} { { \{ 5 \}_i^{(l)} }\over{ [ 4]_i^{(l)}}}
\prod^{N_{l^{\prime}}}_{i=1} { { \{ 7 \}_i^{(l^{\prime})}}\over 
{ [ 8]_i^{(l^{\prime})}}} 
\prod^{N_h}_{i=1} \{ 6 \}_i^{(h)}
\prod^{N_{l^{\prime\prime}}}_{i=1} [6]_i^{(l^{\prime\prime})}  
Q_{ l^{\prime\prime}[N_1 \ldots N_{s^{\prime}}] } 
( x , {\bf x}^{(1)},\ldots , {\bf x}^{(s^{\prime})} ) ,
\eeqcol
\label{llp-bound}
\eeq

respectively. The next process which occurs is $ h+h \to h$ and is 
of third order with respect to the S-matrix 
which means that in the ansatz for $K$ (\ref{K-ansatz}) a 
pole of second order appears. We 
evaluate according to the rules for third order 
forward channel fusion poles and arrive at the following 
recursion relation

\beq
\beqcol 
Q_{hh [N_1 \ldots N_{s^{\prime}}] } 
( x\Omega^2 , x\Omega^{-2} , {\bf x}^{(1)},\ldots , 
{\bf x}^{(s^{\prime})} ) & = & H_{hh} \ x^7 \                  \\
& &                                                            \\
& & {\hspace{-7cm}}  \times
\prod^{N_h}_{i=1} { { \{ 5 \}_i^{(h)} \{ 7 \}_i^{(h)}} \over 
{ [4]_i^{(h)} [8]_i^{(h)} }}  [6]_i^{(h)} 
\prod_{k=l,l^{\prime},l^{\prime\prime}}^{N_k} \{ 6 \}_i^{(k)} \ 
Q_{h [N_1 \ldots N_{s^{\prime}}] } 
( x , {\bf x}^{(1)},\ldots , {\bf x}^{(s^{\prime})} ) . 
\eeqcol
\label{hh-bound}
\eeq

One recognizes that structurally there is no formal difference between 
the recursion relations which arise from a first order pole 
and the one coming from a third order pole respectively. 

\bigskip

Let us now address the problem of the second order pole 
which occurs at $\theta=i \pi /2$. 

The diagram corresponding to this pole and the corresponding 
cut-diagram which leads to a form factor equation is 
as follows (see also \cite{DM2,OOTA}).

\bigskip
\bigskip

{\hspace{6cm}}
{\input{figure3.eps}}

\bigskip
\bigskip

Our goal is to derive a relation for the 
$Q$-polynomials arising from the second order process. 
We take the full form factor (\ref{split}) together 
with our parametrization of the pole part (\ref{K-ansatz}). 
Of course this ansatz can be taken for granted only after 
having established its consistency with the second order 
equation. This means that this ansatz is sufficient 
in the presence of a second order pole if and only if 
we derive a polynomial recursion relation for the $Q$'s. 
This means we do have to verify that the $Q$'s are 
actually polynomials. 

First we have to calculate the momenta of the 
particles in fig.3. These are at the value of the pole 
$ \theta_ l= \theta+ i \pi /2$, $\theta_ h= \theta $, and 
$\theta_{l^{\prime}} = \theta_ {l^{\prime\prime}} = \theta+ i\pi /6$. 

Using the explicit form of the minimal form factors 
\cite{OOTA,MAX} we can deduce the following identitiy in 
$D_4^{(1)}$.  

\beq
\prod_{k=1}^{s^{\prime}} \prod_{i=1}^{N_k} 
{ { F^{\rm{min}}_{l^{\prime}k}( \theta-\theta^{(k)}_i + i \pi /6 ) \ 
  F^{\rm{min}}_{l^{\prime\prime}k}( \theta-\theta^{(k)}_i + i \pi /6 )} 
\over { 
F^{\rm{min}}_{lk}( \theta-\theta^{(k)}_i + i \pi /2 )  \ 
F^{\rm{min}}_{hk}( \theta-\theta^{(k)}_i  ) }} 
= \prod_{i=1}^{N_l} {1\over{ \langle 2 \rangle_{+(\theta_i)}^{(l)} }}\ 
  \prod_{i=1}^{N_h} {1\over{ \langle 1 \rangle_{+(\theta_i)}^{(h)} }} .
\label{fident1}
\eeq

If we then write the equation for the full form factors 
in the following way

\beq
-i \ {\rm res}_{\theta' = \theta+ i \pi /2 } \ 
F_{lh  d_{1}\ldots d_{n}}
(\theta', \theta,\theta_ 1, \ldots, \theta_ n)  =
\Gamma_{l^{\prime\prime}l^{\prime\prime}}^h  
\Gamma_{l^{\prime}l^{\prime\prime}}^l  
F_{l^{\prime} l^{\prime\prime} d_1 \ldots d_n}
(\theta+ i \pi /6, \theta + i \pi /6, \theta_ 1, \ldots, \theta_n) ,
\label{second-order}
\eeq

which is in accordance with the two particle form factor 
equations in an Ising model \cite{DM2} and with \cite{MUSS}, we arrive at 
another polynomial equation for the $Q$'s

\beq
\beqcol
Q_{l h [N_1,\ldots ,N_{s^{\prime}} ] } (x \Omega^2, x \Omega^{-1} , 
{\bf{x}}^{(1)},\ldots , {\bf{x}}^{s^{\prime}} ) & = & 
 x^2 { { \Gamma_{ll}^h  \Gamma_{ll}^l }\over{3\sqrt{3}}} 
{ {F_{l^{\prime}l^{\prime\prime}}^{\rm{min}}(0) } \over 
  {F_{lh}^{\rm{min}}(i \pi /2) }}  
\Omega^{ 2 N_l + N_h}                                        \\
& &                                                              \\
& & {\hspace{-7cm}} \times 
\prod_{i=1}^{N_l} { { \{ 5 \}_i^{(l)} }\over { [4]_i^{(l)} }} 
\prod_{i=1}^{N_h} { { \{ 6 \}_i^{(h)} }\over { [7]_i^{(h)} }} 
\prod_{i=1}^{N_{l^{\prime}}} [6]_i^{(l^{\prime})}
\prod_{i=1}^{N_{l^{\prime\prime}}} [6]_i^{(l^{\prime\prime})}   \    
Q_{l^{\prime} l^{\prime\prime} [N_1,\ldots ,N_{s^{\prime}} ] } 
(x , x  , {\bf{x}}^{(1)},\ldots , {\bf{x}}^{s^{\prime}} ) . 
\eeqcol
\label{Q-second}
\eeq

We can verify the following identity, which should 
be compared with an expression which 
occured in \cite{OOTA} in connection with the second 
order poles in $A_n^{(1)}$

\beq
{ F^{\rm{min}}_{lh}(3 i \pi/6) \over { F^{\rm{min}}_{l h}( 5 i \pi/6)}} 
\ = \ 
{ F^{\rm{min}}_{l^{\prime}l^{\prime\prime}}(0) 
\over { F^{\rm{min}}_{l^{\prime}l^{\prime\prime}}( 4 i \pi/6)}} .
\label{fident2}
\eeq

This identity is necessary in order to guarantee the 
consistency of the solutions of (\ref{Q-second}) with 
the other equations in $D_4^{(1)}$ due to the 
presence of minimal form factors at particular 
values in the equations (\ref{Q-kinemat}) and (\ref{Q-bound}).

It is remarkable to note that even (\ref{second-order}) can more 
or less be treated on the same footing like the first order 
equations for the form factor by realizing that if 
the fictitious vertex (cf. (\ref{vertexdef})) ${\tilde{\Gamma}}_{lh} = 
-i ``{\rm res}``_{\theta= i \pi /2} S_{lh}(\theta )$ is equal 
to the expression $\Gamma_{l^{\prime\prime}l^{\prime\prime}}^h  
\Gamma_{l^{\prime}l^{\prime\prime}}^l  $ which occurs in 
(\ref{second-order}). The quotation marks at the residue 
mean that we have to evaluate at the leading order 
of the singularity.

Moreover (\ref{Q-second}) looks very similar 
to the ones coming from first order processes. In particular 
one might have noticed above that the particles running 
into the local operator acquire 
a factor $ [6] = [h] $ in the recursion coefficient. 
This also happens here where the two light particles 
running into the local opeartor just have this factor in 
the recursion coefficient.

\medskip 

To all the bound state equations the remark on the order 
of the particles which has been made after stating the 
general equation (\ref{Q-bound}) applies. This means that 
we get additional relations for one and the same $Q$-polynomial 
at specific values of the arguments !

\medskip

In the next section we will give some details on the 
calculation on solutions to the recursion relations 
derived above. It turns out that we do neccessarily need 
equation (\ref{Q-second}) which originates from the 
second order pole of the S-matrix. It is needed to 
reduce the degrees of freedom of the solutions. 

This is an important fact because it is known \cite{BCDS1} 
and it was mentioned above 
that the second order diagrams do not really contribute 
to the S-matrix bootstrap but they are needed for 
the corresponding form factor bootstrap !

Let us briefly comment on second order poles in general. 
In principle we can expect an equation for the entire 
form factor of the kind (\ref{second-order}). Even 
though the position of the second order poles of 
the S-matrices in the D-series are known there 
is no closed expression in the literature for the particles 
actually participating in the second order process. 
We have checked for the $D_6$ case that an equation of 
the type (\ref{second-order}) gives a result comparable 
to (\ref{Q-second}) for second order processes. 
However, since (\ref{Q-second}) has a quite nice 
structure one might expect to be able to write 
down a closed recursion formula for all 
admissible second order processes in the D-series.

\bigskip
\bigskip

\section{Some notes on solutions in the $D_4^{(1)}$ case}

\bigskip
\bigskip

Now we are coming to the construction of solutions to the 
recursion relations derived in the previous section. 

It was already pointed out that due to the ansatz for 
the singular part $K$ of the form factor (\ref{K-ansatz}) 
the polynomials $Q_{[N_1,N_2,N_s,N_{s^{\prime}}]}({\bf{x}}^{(1)},
{\bf{x}}^{(2)},{\bf{x}}^{(s)},{\bf{x}}^{(s^{\prime})})$ are 
symmetric at least in each of the components of one 
particular coordinate vector ${\bf{x}}^{(k)}$, with
$k=1,2,s,s^{\prime}$. It might therefore be useful to 
expand $Q$ in symmetric polynomials. A basis in the 
space of these polynomials which is commonly used in 
form factor calculations (see e.g. \cite{ZAM,KM,FMS1}) are 
the so called elementary symmetric polynomials $e^{(n)}_r$. 
However, even though we are going to use them in the 
calculations in this sections we have the 
impression that they might not 
be the appropriate basis for the problem because 
the actual calculations become quite tedious when 
using this basis. We will comment on that problem 
in the final section.

\medskip

Let us introduce some simple facts about 
elementary symmetric polynomials \cite{MAC}. As functions 
of n variables $x_1,x_2, \ldots , x_n$ they are defined 
by the following expression

\beq
\prod_{i=1}^{n} (1+ t\ x_i ) = \sum_{r=0}^{n} e^{(n)}_r \ t^r .
\label{el-symm}
\eeq

Here the superscript on the $e$'s refers to the number of arguments. 
Explicitly 
the $e$'s are given by $e^{(n)}_0=1$, 
$e_1^{(n)} = x_1+x_2 +\ldots x_n$, $\ldots$, 
$e_n^{(n)} = x_1 x_2 \cdots x_n $. 

Comparing (\ref{brack-def}) with (\ref{el-symm}) one recognizes 
at once that all the recursion coefficients in the equations 
for the $Q$-polynomials can in principle be expressed in 
terms of the elementary symmetric polynomials. 

\medskip

The next entity we need is the concept of a partition $\lambda$ 
of a positive integer $m$. For our purposes this is a finite 
sequence of non-negative integers arranged in decreasing 
order $\lambda= ( \lambda_1, \lambda_2,\ldots , \lambda_r)$ 
with $\lambda_1 \ge \lambda_2 \ge \ldots \ge \lambda_r$. 
Characteristic quantities for $\lambda$ are its length 
( $l( \lambda ) =r$ in our example ) and its weight 
$| \lambda | = \sum_{i=1}^r \lambda_i $ (which is equal 
to $m$ in our example). 

Having these concepts we define an elemtary symmetric function 
associated to a partition $\lambda$ by

\beq
E^{(n)}_{\lambda} = e^{(n)}_{\lambda_1} \cdots e^{(n)}_{\lambda_r}.
\label{symm-part}
\eeq

It is then clear that the weight of the partition gives the 
total degree of $E^{(n)}_{\lambda}$ while its length gives 
the maximal power of a particular variable $x_i$ in 
$E^{(n)}_{\lambda}$. The latter is called partial degree. 

Since we have recognized that the coefficients in the 
recursion equations are in fact elementary symmetric polynomials 
we can read off these equations in how much the 
degree of a variable $x_i^{(k)}$ is changed when 
performing one step in the iteration of the 
polynomials $Q$. Since the recursion equations are 
linked it is not quite a straightforward task 
to calculate the partial degree of the coordinates of 
$ ({\bf{x}}^{(1)},{\bf{x}}^{(2)},{\bf{x}}^{(s)}, 
{\bf{x}}^{(s^{\prime})} ) $. Given the polynomial 
$Q_{[N_l,N_{l^{\prime}},N_{l^{\prime\prime}},N_h]}({\bf{x}}^{(l)},
{\bf{x}}^{(l^{\prime})},{\bf{x}}^{(l^{\prime\prime})},{\bf{x}}^{(h)})$ 
where $l,l^{\prime},l^{\prime\prime} \in \{  1,s,s^{\prime} \}$ 
we find for the maximal length of the partitions $\lambda^{(l)}$ 
corresponding to particle $l$

\beq
l(\lambda^{(l)} ) \le 3 ( N_l -1) + 2 (N_{l^{\prime}} +
N_{l^{\prime\prime}} ) + 4 N_h .
\label{l-pardeg}
\eeq

The partial degree of $l^{\prime}$ and $l^{\prime\prime}$ resp. 
can be obtained by just replacing the letters in (\ref{l-pardeg}). 

For the heavy particle we get

\beq
l(\lambda^{(h)} ) \le 7 (N_h -1) + 4(N_l + N_{l^{\prime}} + 
N_{l^{\prime\prime}}) .
\label{h-pardeg}
\eeq

This result reflects again the symmetry of the Dynkin diagram at 
the level of form factors.

\medskip

Due to the presence of four different particles in the theory we 
have to associate an elementary symmetric polynomial to each 
of the particles seperately. We therefore introduce the 
following notation 

\beq 
\Lambda= (\lambda^{(1)} | \lambda^{(2)} | \lambda^{(s)} | 
\lambda^{(s^{\prime})} ) .
\label{total-lambda}
\eeq

There are many different ways to choose initial conditions 
for the symmetric polynomials $Q$, see e.g. \cite{KM}. 
From the degree-formulas (\ref{Q-degree}), (\ref{Q4-degree}) 
one can see that the $Q$-polynomials with only one argument 
are of zero degree which means that they have to be constant. 
We therefore choose the following initial values for the 
$Q$'s

\beq
Q_{[1,0,0,0]}= n_{(1)}, \quad
Q_{[0,1,0,0]}= n_{(2)}, \quad 
Q_{[0,0,1,0]}= n_{(s)}, \quad 
Q_{[0,0,0,1]}= n_{(s^{\prime})} .
\label{initialc}
\eeq

With this data we can compute the two-particle polynomials. 
With $c^{[a,b,c,d]}$ we denote constants which can not be 
determined by the recursion equations.
 
If two light particles are present we find 

\beq
Q_{[2,0,0,0]}( x^{(1)}_1,x^{(1)}_2 ) = 
( {1\over{\sqrt{3}}} H_{ll} n_{(2)} - 3 c^{[2,0,0,0]}) 
E_{ ( 21 | 0 | 0 | 0 ) } + 
c^{[2,0,0,0]}  E_{ ( 1^3 | 0 | 0 | 0 ) } .
\label{2000}
\eeq

We have written down the polynomial for two particles 
of type $1$. The polynomials for particles $s$ and 
$s^{\prime}$ are structurally identical. One has only to 
change the numbering at $Q$ and the $E$'s etc. to the appropriate 
place. 

The next polynomial corresponds to two different light 
particles. We again write only one member of this family, 
the other ones can be obtained by symmetry.

\beq
Q_{[1,0,1,0]} ( x^{(1)}_1,x^{(s)}_1 ) = 
c^{[1,0,1,0]} ( E_{ (1^2 | 0 | 0 | 0 )} + 
                E_{ (0 | 0 | 1^2 | 0 )}  ) + 
( H_{ll^{\prime}} n_{s^{\prime}} + c^{[1,0,1,0]} ) 
E_{ (1 | 0 | 1 | 0 )} .
\label{1010}
\eeq

The polynomial for two heavy particles is already of 
degree 7 and it is on the two particle level not 
possible to eliminate most of the constants. 

\beq
\beqcol
Q_{[0,2,0,0]}( x^{(2)}_1,x^{(2)}_2 ) & = &
c^{[2,0,0,0]}_1 E_{(0 | 1^7 | 0 | 0)} +
c^{[2,0,0,0]}_2 E_{(0 | 2 1^5 | 0 | 0)}     \\

& &  + c^{[2,0,0,0]}_3 E_{(0 | 2^2 1^3 | 0 | 0)} + 
c^{[2,0,0,0]}_4 E_{(0 | 2^3 1 | 0 | 0)} ,    \\
& &                                          \\
& & {\hspace{-3cm}} 
c^{[2,0,0,0]}_1 + c^{[2,0,0,0]}_2 + c^{[2,0,0,0]}_3 + 
c^{[2,0,0,0]}_4 = H_{hh} n_{(2)} .
\eeqcol
\label{0200}
\eeq

The case of one light and one heavy particle is interesting 
because we can see there that the equation arising 
from the second order pole of the S-matrix is needed in 
order to reduce the number of degrees of freedom of the 
solutions. The basis for this particular case which is 
a polynomial of degree 4 is spanned by 
$E_{(1^4 | 0 | 0 | 0)}$, $E_{(1^3 | 1 | 0 | 0)}$, 
$E_{(1^2 | 1^2 | 0 | 0)}$, $E_{(1 | 1^3 | 0 | 0)}$, and 
$E_{(0 | 1^4  | 0 | 0)}$. If we label the corresponding 
constants consecutively for these polynomials we 
see that the first order equations reduce to three 
degrees of freedom (one of them is actually 
reduced by the symmetry properties of the 
$Q$-polynomials at particular arguments mentioned above), 
while the ``second order equations'' 
remove another two degrees of freedom. We then find

\beq
\beqcol 
c^{[1,1,0,0]}_1 &=& {1\over{2}} ( -c^{[1,1,0,0]}_3 +  
                  P (1+ 2 \Omega^{-3} )) , \quad         
c^{[1,1,0,0]}_2  =  ( -{1\over{2}}- { \Omega^{-3}\over{\sqrt{3}}} ) 
                  c^{[1,1,0,0]}_3    
                  - { \Omega^{-3}\over{\sqrt{3}}} (H-P) , \\
& &                                                   \\
c^{[1,1,0,0]}_4 &=& ( -{1\over{2}}- { \Omega^{-3}\over{\sqrt{3}}} ) 
                  c^{[1,1,0,0]}_3    
                  - { \Omega^{-3}\over{\sqrt{3}}}  H 
                  -{1\over{2}} \Omega^{3} P  , \quad              
c^{[1,1,0,0]}_5  = -{1\over{2}} c^{[1,1,0,0]}_3 ,
\eeqcol
\label{1100}
\eeq
where we have set $H=H_{lh} n_{(1)}$ and $P= 3 c^{[0,0,1,1]} + 
H_{ll^{\prime}} n_{(s^{\prime})}$.

It is clear already at this point that the structure of 
the solutions becomes quite complicated already at the 
two particle level due to the high polynomial degrees. 

Of course it is straightforward, however tedious, to compute 
the polynomials for higher particle content as well. 
For example the polynomial $Q_{[2,1,0,0]}$ contains 
almost thirty elementary symmetric polynomials. We do 
prefer not to list the higher cases in this paper. 

We would like to add the following remark. In 
(\ref{l-pardeg}) and (\ref{h-pardeg}) we already 
obtained a result on the partial degree of the 
$Q$-polynomials. It can be taken for granted that 
all the partitions for a specific $Q$ which are 
constrained by the partial degree do contribute 
to the $Q$-polynomial. This means that we will not 
find a case where the coefficients $c^{[a,b,c,d]}_r$ 
for the corresponding basis function do vanish. 
In other words all the elementary symmetric basis 
functions obtained by this general rule do and must 
nonvanishingly contribute to a consistent solution 
$Q_{[a,b,c,d]}$.

\medskip

We do believe that a basis using elementary symmetric 
polynomials is not the appropriate one for explicitly 
calculating the $Q$-part of the form factors and 
therefore would like to postpone a detailed discussion 
of the higher order polynomials. We are going to 
comment a bit more on that point in the next section.

\bigskip
\bigskip

\section{Discussion}

\bigskip
\bigskip

By using the form factor bootstrap and a specific (but 
generic) ansatz (\ref{K-ansatz}) for the singular part of the form factors 
we have been able to derive closed recursion relations 
for $Q$-polynomials within this ansatz. One of these 
equations arises from the kinematical residue equation 
of the form factor bootstrap. The other one which 
is at first a consequence of the first order bound 
state residue equation of the form factor bootstrap 
turned out with some slight modifications to be 
applicable for the fusings in $D_n^{(1)}$-ATFT which 
come from third order poles of the S-matrices as well 
which is a consequence of the S-matrix bootstrap.

We did comment on the consequences of the second 
and fourth order poles of the S-matrix on the 
$Q$-polynomials. Since at present there is, 
in contrast to the $A_n^{(1)}$ cases, no rule saying 
which particles do participate in the Feynman 
diagrams contributing to the even order poles 
of the S-matrix we cannot give a general closed equation 
for the $Q$-polynomials in these cases. However, 
it is clear that at least some of them are needed 
in a particular theory in order to reduce 
the number of degrees of freedom of the $Q$'s.

We treated $D_4^{(1)}$-ATFT as a special case of 
our general analysis. We have shown how the 
symmetries of the Dynkin diagram are in a sense 
maintained in the equations for the $Q$-polynomials 
in this case. Moreover we treated the influence of 
the second order poles of the S-matrix on the 
form factors in detail and derived another equation for the 
$Q$-polynomials. 

In attempting to calculate solutions in the 
$D_4^{(1)}$ case we realized that due to the 
high polynomial degree of $Q$ the solution 
spaces are very complicated when a basis 
for $Q$ consisting of elementary symmetric 
polynomials is used.

In order to find solutions without going through 
tedious computations it would first be interesting 
to find a Lie algebraic interpretation of the 
recursion relations. Second one has to think of 
another polynomial basis in the solution spaces. 
A direct guess would be to use Macdonald symmetric 
polynomials \cite{MAC}. This is because these 
polynomials are symmetric polynomials which 
depend on two parameters. This fits well into 
the affine Toda case since we do have exactly 
two parameters in the recursion equations. One of 
them is $\Omega= e^{i \pi /h}$ and for the other 
one we can take $\Omega^B$ with $B$ being the 
effective coupling (\ref{Bdef}). Due to the 
weak-strong duality $B \to 2-B$ \cite{BCDS1} and 
the symmetry of the $Q$-polynomials which was pointed 
out in the discussion after equation (\ref{Q-bound}) 
we have some hint that the Macdonald polynomials 
might be a proper candidate (see \cite{MAC}). Also their connection 
to root systems of Lie algebra (see e.g. \cite{KV}) 
might be of use for the present problem.

Let us make one last comment on the deformation aspect 
of the solution spaces. In \cite{ZAM} form factors 
of the scaling Lee-Yang model were calculated. If 
one considers the closed solution of \cite{ZAM} for 
the Q-polynomials one finds that a solution to 
the n particle case is exactly one skew Schur function 
\cite{MAC} 

\beq
Q_n \sim s_{\lambda /  \mu } (x_1,\ldots , x_n) , \qquad 
\lambda^{\prime}= (2n-2, 2n-3, \ldots , n-2), \quad 
\mu^{\prime} = (2n-4, 2n-6, \ldots , 0),
\eeq

where $\lambda^{\prime},\mu^{\prime}$ are the conjugate 
partitions. 

The scaling Lee-Yang model can in some sense be considered 
as a special case of the sinh-Gordon model (more or less by 
setting $B \to 0$). Or the latter one 
is a deformation of the first. In \cite{KM} it was shown 
that for the sinh-Gordon form factor problem it is possible 
to find a closed solution for the $Q$-polynomials as well. 
The result is 

\beq
Q_n(k) = {\rm det} M_{ij}(k), \quad M_{ij} (k) = [i-j+k]
e^{(n)}_{2i-j} ,
\eeq
where M is a $(n-1) \times (n-1)$ matrix and 
$[n]= {\rm sin}(n B/2)/ {\rm sin}(B/2)$. This solution 
is a then a certain deformation of the skew Schur function 
above. 

It would be interesting in the case of $D_n^{(1)}$ 
theories if in certain limits models exist for which a closed 
solution for the $Q$'s can be constructed easily which 
then would allow for hints for the generic cases.

\bigskip
\bigskip
\bigskip
\bigskip

{\bf Acknowledgements}

The author is grateful to Professor R. Sasaki for many helpful 
and stimulating discussions and for advise concerning 
the higher order pole properties of the form factors, and to T. Oota 
for sharing his knowledge 
on form factors in the $A_n^{(1)}$ cases. Conversations 
with M. Niedermaier, P. Dorey, R. Tateo, S. Pratik Khastgir, and 
H.-C. Fu have been of much help as well.

This work was supported by a fellowship of the Japan Society 
for the Promotion of Science (JSPS) and also partially by the 
Alexander von Humboldt-Gesellschaft.


\bigskip
\bigskip




\begin{thebibliography}{69}
%
\bibitem{SMIRNOV} F.A. Smirnov: {\it Form factors in completely 
integrable models of quantum field theory}, Adv. Series in 
Math. Phys. 14. (World Scientific, Singapore, 1992).
%
\bibitem{KW} M. Karowski, P. Weisz: {\it Nucl. Phys. B} 
{\bf 139} (1978) 455.
%
\bibitem{CM} J.L. Cardy, G. Mussardo: {\it Nucl. Phys. B} 
{\bf 340} (1990) 387.
%
\bibitem{YZ} V.P. Yurov, Al.Z. Zamolodchikov: 
{\it Int. J. Mod. Phys. A} {\bf 6} (1991) 3419
%
\bibitem{ZAM} Al.Z. Zamolodchikov: {\it Nucl. Phys. B} {\bf 348} 
(1991) 619.
%
\bibitem{KM} A. Koubek, G. Mussardo: {\it Phys. Lett. B} {\bf 311} 
(1993) 193.
%
\bibitem{FMS1} A. Fring, G. Mussardo, P. Simonetti: 
{\it Nucl. Phys. B} {\bf 393} (1993) 413.
%
\bibitem{FMS2} A. Fring, G. Mussardo, P. Simonetti: 
{\it Phys. Lett. B} {\bf 307} (1993) 83.
%
\bibitem{DM} G. Delfino and G. Mussardo: {\it Phys. Lett. B} 
{\bf 324} (1994) 40.
%
\bibitem{KOUB} A. Koubek: {\it Nucl. Phys. B} {\bf 428} (1994) 655.
%
\bibitem{DM2} G. Delfino, G.Mussardo: {\it Nucl. Phys. B} {\bf 455} 
(1995) 724.
%
\bibitem{OOTA} T. Oota: {\it Nucl. Phys. B} {\bf 466} (1996) 361. 
%
\bibitem{KAC} V. Kac: {\it Infinite dimensional Lie algebras} 
3rd edition, Cambridge Univ. Press, Cambridge 1990.
%
\bibitem{BS} H.W. Braden, R. Sasaki: {\it Phys. Lett. B} 
{\bf 255} (1991) 343.
%
\bibitem{BCDS1} H.W. Braden, E. Corrigan, P.E. Dorey, R. Sasaki: 
{\it Nucl. Phys. B} {\bf 338} (1990) 689. 
%
\bibitem{BCDS2} H.W. Braden, E. Corrigan, P.E. Dorey, R. Sasaki: 
{\it Nucl. Phys. B} {\bf 356} (1991) 469.
%
\bibitem{DOR} P. Dorey: {\it Nucl. Phys. B} {\bf 358} (1991) 654, 
{\it Nucl. Phys. B} {\bf 374} (1992) 741.
%
\bibitem{DDV} C. Destri, H.J. De Vega: {\it Nucl. Phys. B} 
{\bf 358} (1991) 251.
%
\bibitem{MAX} M.R. Niedermaier: {\it Nucl. Phys. B} {\bf 424} 
(1994) 184.
%
\bibitem{OOTA2} T. Oota: private communication. 
%
\bibitem{MUSS} C. Acerbi, G. Mussardo, A. Valeriani: 
preprint SISSA-105-96-EP, hep-th/9609080.
%
\bibitem{MAC} I.G. MacDonald: {\it Symmetric functions and 
Hall polynomials} 2nd edition, Clarendon Press, Oxford 1995.
%
\bibitem{KV} N.Ja. Vilenkin, A.U. Klimyk: {\it Representations 
of Lie Groups and Special Functions, Recent Advances}, Kluwer, 
Dordrecht 1995.
%
\end{thebibliography}
\end{document}